\documentclass[aps,prl,twocolumn,superscriptaddress,showpacs]{revtex4-1}

\usepackage{amsmath}
\usepackage{graphicx}
\usepackage{calc}
\usepackage{bm}

\usepackage[T2A]{fontenc}
\usepackage[cp1251]{inputenc}
\usepackage[english]{babel}

\begin{document}

\title{ Magnetocaloric effect for the topological semimetal Co$_3$Sn$_2$S$_2$ due to the antiferromagnetic coupling of the bulk and surface spin-polarized phases}

\author{N.N. Orlova}
\author{V.D. Esin}
\author{A.V. Timonina}
\author{N.N. Kolesnikov}
\author{E.V. Deviatov}

\affiliation{Institute of Solid State Physics of the Russian Academy of Sciences, Chernogolovka, Moscow District, 2 Academician Ossipyan str., 142432 Russia}

\date{\today}

\begin{abstract}
We experimentally investigate magnetocaloric effect for the topological magnetic Weyl semimetal  Co$_3$Sn$_2$S$_2$ in a wide temperature range. The isothermal magnetic entropy change $\Delta S$ is calculated from the experimental magnetization curves by using Maxwell relation. In addition to the expected $\Delta S$ peak at the Curie temperature $T_C$, we obtain another one at the temperature $T_{inv}$ of the hysteresis inversion, which is the main experimental result. The inverted hysteresis usually originates from the antiferromagnetic coupling between two  magnetic phases. For Co$_3$Sn$_2$S$_2$ topological magnetic Weyl semimetal these phases are the ferromagnetic bulk and the spin-polarized topological surface states. Thus, the pronounced magnetocaloric effect at $T_{inv}$ is determined by the bulk magnetization switching by the exchange bias field of the surface spin-polarizad phase, in contrast to the ferromagnetic-paramagnetic transition at the Curie temperature $T_C$. For possible applications of magnetocaloric effect,  Weyl semimetals open a new way to shift from ferromagnetic to the antiferromagnetic systems without loss of efficiency, but with higher reversibility and with smaller energy costs. 
\end{abstract}


\maketitle

\section{Introduction}

The nontrivial properties of topological semimetals originate from the existence of distinct band-touching Dirac points in the bulk electronic spectrum. Weyl semimetals  are also characterized by broken inversion or time-reversal symmetries, so every Dirac point is  splitted on two Weyl nodes with opposite chiralities~\cite{armitage}.  As a result,  Fermi-arc surface states are formed as the arcs between the projections of Weyl nodes on the surface Brillouin zone~\cite{armitage,kagome_arcs}. 

Among the magnetic Weyl semimetals, kagome-lattice ferromagnet Co$_3$Sn$_2$S$_2$ is a typical example hosting spin-polarized Fermi arc surface states~\cite{asymmr}. The giant anomalous Hall effect has been reported~\cite{kagome,kagome1} for Co$_3$Sn$_2$S$_2$, as a sign of a magnetic Weyl phase~\cite{armitage}. Fermi arcs were directly visualized by scanning tunneling spectroscopy~\cite{kagome_arcs}.  Spin-polarized surface textures have also been experimentally shown~\cite{CoSnS_skyrmion} for Co$_3$Sn$_2$S$_2$   as well as for other magnetic topological semimetals~\cite{CrGeTe, FGT_skyrmion}, as a result of the spin-momentum locking in the  topological Fermi-arc surface states~\cite{armitage}.

For Co$_3$Sn$_2$S$_2$, usual rectangular ferromagnetic hysteresis is changed to the inverted one~\cite{Orlova} at the 140~K transition temperature, demonstrating the presence of the second magnetic phase~\cite{invhyst}. Antiferromagnetic coupling between the  magnetic phases provides  the exchange bias field, which facilitate the magnetization switching~\cite{Exb,Invh,Eb_Invh_oxfilm}. While the first phase is obviously the ferromagnetic bulk, the surface spin textures should be regarded as the second magnetic phase in  magnetic topological semimetals. Due to the topological protection, the surface magnetic phase shows excellent temperature stability, as it has been confirmed by first order reversal curve analysis and the bow-tie~\cite{bow-tie} hysteresis loops at high temperatures~\cite{Orlova}. The described two-phase behavior is quite universal for different magnetic topological semimetals, e.g. for Co$_3$Sn$_2$S$_2$ and Fe$_3$GeTe$_2$, only the characteristic temperatures differ for these materials~\cite{Orlova}.
 
For the ferromagnets, the magnetocaloric effect (MCE) is of special interest due to the possible applications~\cite{MCEmater1,MCEmater2}. It is expected that the MCE-based refrigerators  will be energetically more efficient than the usual ones~\cite{ref devices}. Thus, different MCE variations are of extensive investigation nowadays, e.g. multicalloric effect~\cite{multical1,multical2,multical3},  the giant magnetocaloric effect~\cite{GMCE1,GMCE2}, and the rotating magnetocaloric effect~\cite{deltaS1/RMCE}. For the topological Weyl semimetal Co$_3$Sn$_2$S$_2$, usual magnetocaloric effect has been demonstrated around the Curie temperature~\cite{deltaS2,deltaS3}, as well as  the rotating magnetocaloric effect~\cite{deltaS1/RMCE} due to the large magnetic anisotropy~\cite{deltaS1/RMCE}. 

The magnetocaloric effect appears as the entropy change $\Delta S$ while the system undergoes the transition between differently ordered state, usually, ferromagnetic-paramagnetic  transition at the Curie point. On the other hand, the bulk magnetization switching by the exchange bias field of the surface spin-polarizad phase~\cite{Orlova} should also be accompanied by  the magnetic entropy change $\Delta S$ in magnetic topological semimetals. In other words, surface-induced magnetocaloric effect can be expected for magnetic topological semimetals, which is worth the experimental investigation. For possible applications of the magnetocaloric effect, utilizing of antiferromagnetic coupling between the surface and bulk phases should increase energy efficiency of a refrigerator.  

Here, we experimentally investigate magnetocaloric effect for the topological magnetic Weyl semimetal  Co$_3$Sn$_2$S$_2$ in a wide temperature range. The isothermal magnetic  entropy change $\Delta S$ is calculated from the experimental magnetization curves by using Maxwell relation. In addition to the expected $\Delta S$ peak at the Curie temperature $T_C$, we obtain another one at the temperature $T_{inv}$ of the hysteresis inversion, which is the main experimental result.

\section{Samples and techniques}

 Co$_3$Sn$_2$S$_2$ single crystals were grown by the gradient freezing method. Initial load of high-purity elements taken in stoichiometric ratio was slowly heated up to $920^{\circ}$C in the horizontally positioned evacuated silica ampule, held for 20 h and then cooled with the furnace to the ambient temperature at the rate of 20 degree/h. The obtained ingot was cleaved in the middle part. Electron probe microanalysis of cleaved surfaces and X-ray diffractometry of powdered samples confirmed stoichiometric composition and the space group $R\overline{3}m$ (No. 166) of the crystal. Since the Laue patterns confirm the hexagonal structure with (0001) as cleavage plane, we use small Co$_3$Sn$_2$S$_2$ flakes, which are obtained by a mechanical cleavage from the initial single crystal.

The experimental magnetization curves have been obtained by the Lake Shore Cryotronics 8604 VSM magnetometer. It is equipped with nitrogen flow cryostat for measurements from the room temperature down to 80~K, while the  Curie temperature is known~\cite{Tc1,Tc2} to be about 173~K for Co$_3$Sn$_2$S$_2$ ferromagnet. A flake is mounted to the magnetometer sample holder by a low temperature grease, which has been tested to have a negligible magnetic response. The sample can be rotated in magnetic field, so the hard and easy magnetization directions (i.e. sample orientation)  can be obtained from angle-dependent magnetization. 

We investigate the magnetocaloric effect as the isothermal magnetic  entropy change $\Delta S$ which is calculated from the experimental magnetization curves by using Maxwell relation~~\cite{DeltaS,technique}:
 
 \begin{equation}
\Delta S = \mu_0\int^H_0 \left(\frac{\partial M}{\partial T}\right)_H dH \label{DeltaS}
\end{equation}

As a first step, a set of $M(H,T=const)$ isotherms is obtained for the required magnetic field range. Subsequently, the first derivative $\left(\frac{\partial M}{\partial T}\right)_H$ is calculated as a result of subtraction $\left(\frac{\Delta M}{\Delta T}\right)(H,T)$ of every two neighbor curves. As a final step, $\Delta S$ is achieved for every temperature point by  integration over the magnetic field range~\cite{technique}. 

Before any set of  $M(H,T=const)$ isotherms, it is necessary to set up the stable initial magnetic state of the sample. In our experiment, it is performed by cooling the sample in zero magnetic field (ZFC protocol) from the room temperature to 80~K one, and subsequent sample magnetization along the easy axis by sweeping the field between -15~kOe and +15~kOe values. Afterward, the external field is switched to zero, so the temperature is always stabilized at zero field for every $M(H,T=const)$ isotherm.

\section{Experimental results}

\begin{figure}
\includegraphics[width=1\columnwidth]{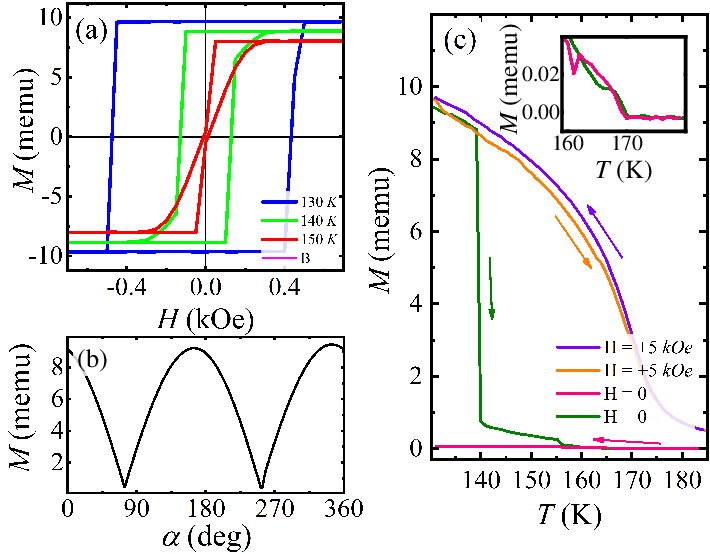}
\caption{(Color online) (a) Examples of hysteresis loops for a 1.42~mg Co$_3$Sn$_2$S$_2$ flake for different temperatures around the $T_{inv}$. The loops are of strictly rectangular shape with step-like magnetization switchings at 130~K and below, which confirms the single domain magnetic state of the Co$_3$Sn$_2$S$_2$ flake. At 140~K, the slanted hysteresis sections appear, so the loop is of bow-tie type~\cite{bow-tie}. At 150~K and for high temperatures, step-like magnetization swichings occur before the magnetic field inversion, which is known as inverted hysteresis~\cite{Exb,Invh,Eb_Invh_oxfilm}. The curves are obtained after zero-field cooling, the magnetic field is along the Co$_3$Sn$_2$S$_2$ easy axis direction. (b) Angle dependence of  magnetization $M$ for the Co$_3$Sn$_2$S$_2$ single crystal flake. The curve is obtained at 5~kOe magnetic field, i.e. for fully saturated sample magnetization. (c) Temperature-dependent sample magnetization $M(T)$. Green and pink curves are obtained in zero magnetic field after zero-field cooling and the initial sample magnetization, see the Samples section. While increasing the temperature,  the inverted hysteresis transition appears as a sharp drop of magnetization at T$_{inv}$ =140~K. The backward zero-field cooling $M(T)$ curve shows negligible magnetization value. For the orange and purple $M(T)$ curves, both cooling and heating are performed at 5~kOe magnetic field (FC/FH protocols). The curves are smooth, without any peculiarity at  
T$_{inv}=140$~K, and with negligible difference for the FC and FH curves. Inset shows the enlarged region around the Curie temperature $T_C$ = 173--175~K for the zero-field cooling and heating curves. }
\label{hysteresis}
\end{figure}

Fig.~\ref{hysteresis} (a) shows  hysteresis loops for a 1.42~mg  Co$_3$Sn$_2$S$_2$ flake at three different temperatures, 130~K, 140~K, and 150~K, respectively. Because of the excellent reproducibility of the magnetization results from different Co$_3$Sn$_2$S$_2$ flakes~\cite{Orlova}, below we present the data from this sample only.  The curves are obtained after zero-field cooling, the magnetic field is normal to the (0001) cleavage plane, i.e. it is directed along the Co$_3$Sn$_2$S$_2$ easy axis. The sample orientation is verified by the angle dependence of magnetization in Fig.~\ref{hysteresis} (b). 

Hysteresis loops are of strictly rectangular shape with step-like magnetization switchings below 130~K, which confirms the single domain magnetic state of bulk Co$_3$Sn$_2$S$_2$, see Fig.~\ref{hysteresis} (a), and Ref.~\cite{Orlova} for details. At higher temperatures, coercivity is diminishing, the slanted sections appear, so one can see a bow-tie~\cite{bow-tie}  hysteresis loop at 140~K in Fig.~\ref{hysteresis} (a). Above 150~K, step-like magnetization swichings  occur before the magnetic field inversion, which is known as inverted hysteresis~\cite{Exb,Invh,Eb_Invh_oxfilm}. The inverted hysteresis originates from  antiferromagnetic coupling between the  magnetic phases~\cite{Exb,Invh,Eb_Invh_oxfilm}, which provides  the exchange bias field. For Co$_3$Sn$_2$S$_2$, two phases has been directly demonstrated by FORC analysis~\cite{Orlova}. 

Fig.~\ref{hysteresis} (c) shows temperature-dependent magnetization as two pairs of $M(T)$ curves.  One  pair is obtained in zero magnetic field after zero-field cooling and the initial sample magnetization, as it is described in the Samples section. While increasing the temperature,  the inverted hysteresis transition appears as a sharp drop of magnetization at transition temperature T$_{inv}$ =140~K,  see Fig.~\ref{hysteresis} (c). At higher temperatures,  the transition to the paramagnetic state can be identified  at  $T_C$ = 173--175~K in the inset to Fig.~\ref{hysteresis} (c), which well corresponds to the known Co$_3$Sn$_2$S$_2$ Curie temperature~\cite{Tc1,Tc2}. The backward zero-field cooling $M(T)$ curve shows negligible magnetization value, as it should be expected for a ferromagnet while going from above the Curie temperature.  For the second  pair of $M(T)$ curves, both cooling and heating (FC/FH protocols) are performed at 5~kOe magnetic field. For both FC/FH curves, magnetization drops to zero at the Curie temperature $T_C$. In the FC case, 5~kOe magnetic field is applied at 190~K~$>T_C$ temperature  along the easy magnetization axis, with subsequent cooling to 130~K. Since 5~kOe magnetic field  is well above the inverted hysteresis region in  Fig.~\ref{hysteresis} (a), the curves are smooth, without any peculiarity at  T$_{inv}$ =140~K, and with negligible difference for the FC and FH curves.

\begin{figure}[t]
\center{\includegraphics[width=\columnwidth]{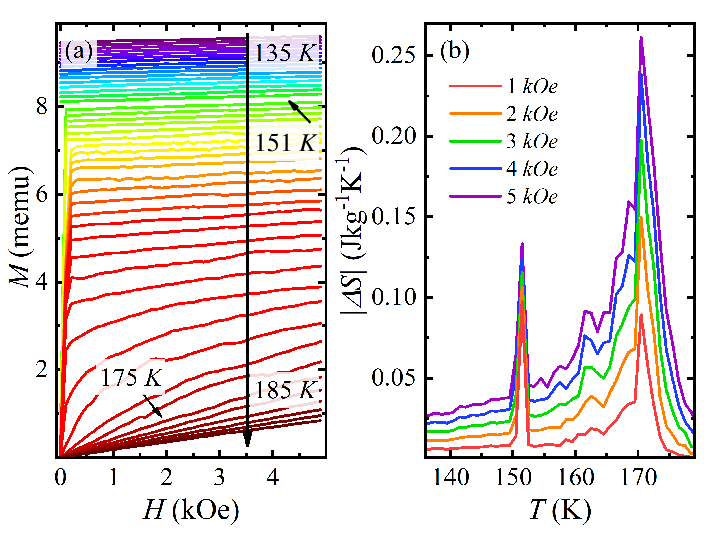}}
\caption{(Color online) (a) $M(H,T=const)$ isothermal magnetization curves in a wide temperature range from 135~K to 185~K with 1~K step. The initial sample state is obtained after ZFC from room temperature to 80~K with subsequent sample magnetization, see the Samples section. (b) 
The absolute value of the isothermal magnetic entropy change $|\Delta S|$ as obtained from Eq.(\ref{DeltaS}),  for different limits of the  magnetic field $H$. In addition to the known~\cite{deltaS1/RMCE,deltaS2,deltaS3} $|\Delta S|$ peak at the Curie temperature $T_C$, we obtain another one at the inversion point $T_{inv}$, which is the main experimental result. The peaks' amplitudes are comparable at both temperatures, $T_C$ and $T_{inv}$.} 
\label{Isoterms}
\end{figure}

For the magnetocaloric effect investigation, $M(H,T=const)$ isothermal magnetization curves are obtained in a wide temperature range, to cover both the Curie temperature $T_C$ and the inversion transition $T_{inv}$, see Fig.~\ref{Isoterms} (a). Since magnetocaloric effect is well studied near the Curie temperature~\cite{deltaS2,deltaS3}, it can be regarded as a reference to confirm the correctness of our procedure. The initial sample state is obtained after ZFC from room temperature to 80~K with subsequent magnetization in the external field up to $\pm$15~kOe, as described above. Afterward, temperature is stabilized in zero field with 1~K step, $M(H)$ curves are obtained from 135~K to 185~K, see Fig.~\ref{Isoterms} (a). 

The isothermal magnetic entropy change $\Delta S$ can be calculated by Eq.(\ref{DeltaS}) for different limits of the  magnetic field $H$, absolute $|\Delta S|$ values are presented in Fig.~\ref{Isoterms} (b). For the widest field range (up to 5~kOe), the obtained $|\Delta S|$ behavior well correspond to the known~\cite{deltaS1/RMCE,deltaS2,deltaS3} for  Co$_3$Sn$_2$S$_2$:  Fig.~\ref{Isoterms} (b) shows $|\Delta S(T)| = 0.26$~J$\times$kg$^{-1}\times$K$^{-1}$ around the Curie temperature $T_C$, similarly to Ref.~\cite{deltaS3}. The $|\Delta S|$ peak value is linearly diminishing for narrower field ranges, while the qualitative behavior is the same, see Fig.~\ref{Isoterms} (b).   

As the main experimental result, Fig.~\ref{Isoterms} (b) shows narrow $|\Delta S|$ peak at the inversion temperature $T_{inv}$, in addition to the $|\Delta S|$ peak around the Curie point. The peaks' amplitudes are comparable at both temperatures, $T_C$ and $T_{inv}$.  Due to the abrupt changes in $M(H,T)$ at the $T_{inv}$ transition point in Fig.~\ref{hysteresis} (a) and (c), the detailed measurements are required with smaller temperature step to avoid the $|\Delta S|$ peak smoothing around the $T_{inv}$.

\begin{figure}[t]
\center{\includegraphics[width=\columnwidth]{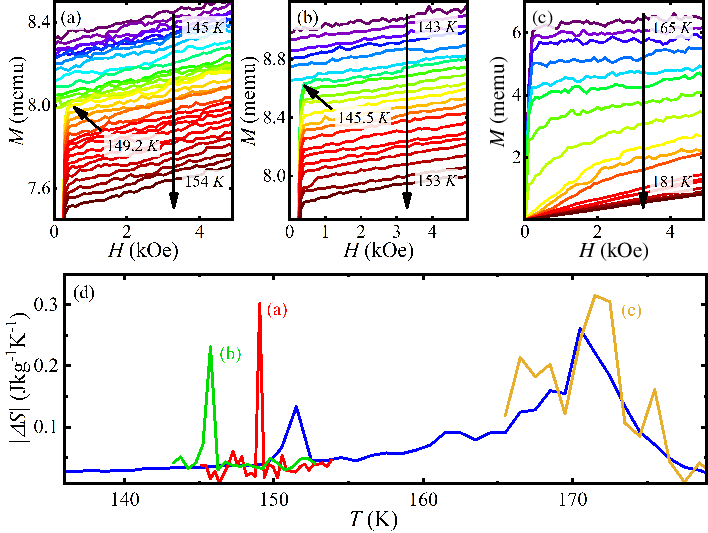}}
\caption{(Color online) (a) $M(H,T=const)$ isothermal magnetization curves for the narrow 145 -- 155~K temperature interval with 0.3~K step, ZFC protocol. The curves are obtained after initial ZFC with subsequent magnetization, similarly to ones in Fig.~\ref{Isoterms} (a). (b) $M(H,T=const)$ isothermal magnetization curves for the 143 -- 153~K temperature interval with 0.5~K step. The curves are obtained after the initial cooling at 5~kOe magnetic field (FC protocol). (c) $M(H,T=const)$ isothermal magnetization curves around the Curie temperature $T_C$ after the field cooling at 5~kOe. (d) The calculated absolute value of the isothermal magnetic entropy change $|\Delta S|$. The red, green and brown curves are  obtained from the  $M(H,T=const)$ data, presented in (a), (b) and (c),  respectively. The blue curve is a reference, it is reproduced from Fig.~\ref{Isoterms} (b) for the same 5~kOe field range. The magnetocaloric effect is practically insensitive to the initial sample state around the Curie point $T_C$ = 173--175~K, while the $|\Delta S|$ peak position depends on the the cooling procedure for inverted hysteresis transition. From the peaks' amplitudes, the magnetocaloric effect is as well efficient for inverted hysteresis transition, as  for the usual ferromagnetic-paramagnetic transition at $T_C$.
}
\label{Isoterms3sets}
\end{figure}

Fig.~\ref{Isoterms3sets} (a) shows  $M(H,T=const)$ isothermal magnetization curves for the narrow 145 -- 155~K temperature interval with 0.3~K step. The curves are obtained after initial ZFC with subsequent magnetization, similarly to ones in Fig.~\ref{Isoterms} (a). The calculated  $|\Delta S|$ is depicted in Fig.~\ref{Isoterms3sets} (d) as a red curve for the 5~kOe magnetic field range, while the reference blue one is from Fig.~\ref{Isoterms} (b). The $|\Delta S|$ peak at $T_{inv}$ is even narrower for the smaller temperature step, the peak height is of the same  $|\Delta S(T)| \approx 0.3$~J$\times$kg$^{-1}\times$K$^{-1}$ value as at the Curie temperature $T_C$. Thus, the magnetocaloric effect is of the same value for the ferromagnetic-paramagnetic transition at $T_C$ and for inverted hysteresis transition at much lower temperature $T_{inv}$. 

However, the position of the inversion point $T_{inv}$ varies from cooling to cooling, as it can be seen from red and blue curves in Fig.~\ref{Isoterms3sets} (d). Being determined by interaction of two magnetic phases, $T_{inv}$ is mostly affected by the bulk coercivity value, since the topological surface phase is of excellent stability~\cite{Orlova}. For a particular Co$_3$Sn$_2$S$_2$ single crystal with definite shape and size, the bulk coercivity  varies due to the stress and defects redistribution in different cooling cycles, leading to $T_{inv}$ within 140--151~K temperature interval. 

Due to the observed $T_{inv}$ variation, the stability of the obtained results for $|\Delta S|$ should be demonstrated by measurements with different, e.g. field-cooling  protocol.  Fig.~\ref{Isoterms3sets} (b) shows  $M(H,T=const)$ isothermal magnetization curves for the 143 -- 153~K temperature interval with 0.5~K step. The curves are obtained after the initial cooling at 5~kOe magnetic field, the calculated $|\Delta S|$ is depicted in Fig.~\ref{Isoterms3sets} (d) as a green curve. The result is qualitatively similar to the ZFC case: the $|\Delta S|$ peak is of approximately same height, it is wider due to the increased temperature step, while the peak position is considerably changed since coercivity is always smaller in the FC case. 

As a reference, we reproduce $M(H,T=const)$ isothermal magnetization curves around the Curie temperature $T_C$ after the field cooling at 5~kOe, see Fig.~\ref{Isoterms3sets} (c). The result of the $|\Delta S|$ calculation is shown in Fig.~\ref{Isoterms3sets} (d) as a brown curve. The $|\Delta S|$ is of the same height as for the ZFC case, it is centered at the same $T_C$ value. Thus, we should conclude that the magnetocaloric effect is practically insensitive to the cooling procedure and, therefore, the initial sample state around the Curie point $T_C$ = 173--175~K.

\section{Discussion}

As a result, in addition to the ferromagnetic-paramagnetic transition at the Curie temperature $T_C$,   the pronounced  magnetocaloric effect is observed at much lower temperature $T_{inv}$, 
where the effect of the second magnetic phase appears as the inverted hysteresis loop.  The isothermal magnetic entropy $\Delta S$  at $T_{inv}$ is of the same  $\approx 0.3$~J$\times$kg$^{-1}\times$K$^{-1}$ value as at the Curie temperature $T_C$.

The corresponding  temperature variation $\Delta T$ can be estimated~\cite{DeltaTad,DeltaS}  from $\Delta S(T)$ as:

\begin{equation}\label{DeltaT}
\Delta T = -\frac{T\Delta S}{C_p(H)}
\end{equation} 
  
The Co$_3$Sn$_2$S$_2$ heat capacity $C_p$ equals~\cite{Cp} to 286~J$\times$kg$^{-1}\times$K$^{-1}$ and 307~J$\times$kg$^{-1}\times$K$^{-1}$  for 147~K and 173~K, respectively. Thus, one can estimate $\Delta T \approx 0.15 - 0.17$~K for these transition points. 

In principle, coexistence of two magnetic phases can be anticipated for Co$_3$Sn$_2$S$_2$ from $M(T)$ measurements in fixed magnetic fields~\cite{CoSnS_Tdepend1, CoSnS_Tdepend2, CoSnS_spin glass} and from the AHE hysteresis~\cite{CoSnS_spin glass}. While the first phase is obviously the ferromagnetic bulk, the existence of the second phase was explained by disorder effects~\cite{CoSnS_Tdepend1} or even by the spin-glass state~\cite{CoSnS_spin glass}.   However,  FORC measurements~\cite{Orlova} directly demonstrate multi-phase behavior even at lowest temperatures, with excellent temperature stability of the second phase in comparison with the main ferromagnetic one~\cite{Orlova}. Also, qualitatively similar behavior is observed for Co$_3$Sn$_2$S$_2$ samples of extremely different size (two orders of magnitude in Ref.~\cite{Orlova}), and for another magnetic topological semimetal FGT, demonstrating  universal character of the second phase~\cite{Orlova}.  Thus, it can hardly be ascribed to temperature-induced disorder effects~\cite{CoSnS_Tdepend1, CoSnS_Tdepend2, CoSnS_spin glass}.

On the other hand,  Co$_3$Sn$_2$S$_2$ is the magnetic semimetal with topologically protected Fermi-arc surface states~\cite{kagome_arcs,asymmr,CoSnS_Ws1,CoSnS_Ws2}. Due to the spin-momentum locking, one can expect  skyrmion-like topological surface spin textures, as it has been confirmed experimentally for Co$_3$Sn$_2$S$_2$ magnetic Weyl semimetal~\cite{CoSnS_skyrmion}.  Thus, one should ascribe the second, temperature-stable magnetic phase to the surface states in topological semimetals, since the surface spin textures are inherent for magnetic topological semimetals due to the spin-momentum locking in the  topological surface states\cite{Orlova}. 
Also, the  hysteresis loops are of the bow-tie type in Fig.~\ref{hysteresis} (a), which is usually ascribed to surface spin textures~\cite{Pt/Co/Ta,Pt/Co/Ta1,Pt/Co/Ta2,Co/Pd,Ir/Fe/Co/Pt}. 

In this case, the magnetocaloric effect at $T_{inv}$ is determined by the transition from the ferromagnetic bulk spin ordering to the antiferromagnetic interaction of the bulk and surface spin-polarized phases. The exchange field of the surface phase is strong enough to induce magnetization direction switching of the bulk ferromagnetic phase. In other words, isothermal magnetic entropy $\Delta S$ is determined by the reorientation of the magnetization direction for the bulk ferromagnetic phase at $T_{inv}$, while $\Delta S$ reflects the transition from ferromagnetic to the paramagnetic ordering at the Curie temperature $T_C$. Similar values of  $\Delta S$ and, therefore,  $\Delta T$ at both transition points in Fig.~\ref{Isoterms3sets} (d) indicate strong influence of the surface state induced spin polarization on the bulk ferromagnetic magnetization in Weyl topological semimetals. 

For possible applications of magnetocaloric effect, Weyl semimetals open a new way to transfer from ferromagnetic to the antiferromagnetic systems without loss of efficiency in $\Delta S$ and, therefore, in  $\Delta T$, but with higher reversibility and with smaller energy costs, which is confirmed by much narrower $\Delta S$ peak at $T_{inv}$ in Fig.~\ref{Isoterms3sets} (d).

\section{Conclusion}
As a conclusion, we experimentally investigate magnetocaloric effect for the topological magnetic Weyl semimetal  Co$_3$Sn$_2$S$_2$ in a wide temperature range. The isothermal magnetic entropy change $\Delta S$ is calculated from the experimental magnetization curves by using Maxwell relation. In addition to the expected $\Delta S$ peak at the Curie temperature $T_C$, we obtain another one at the temperature $T_{inv}$ of the hysteresis inversion, which is the main experimental result. The $|\Delta S|$ peak  height is of the same  $\approx 0.3$~J$\times$kg$^{-1}\times$K$^{-1}$ value as at the Curie temperature $T_C$. Thus, the magnetocaloric effect is of the same value for the ferromagnetic-paramagnetic transition at $T_C$ and for inverted hysteresis transition at much lower temperatures. Similar values of  $\Delta S$ and, therefore,  $\Delta T$ at both transition points in Fig.~\ref{Isoterms3sets} (d) indicate strong influence of the surface state induced spin polarization on the bulk ferromagnetic magnetization in Weyl topological semimetals.

\section{Acknowledgement}
We wish to thank S.S~Khasanov for X-ray sample characterization.


\begin{thebibliography}{99}

\bibitem{armitage} N.P.~Armitage, E.J.~Mele, and A.~Vishwanath,  Rev. Mod. Phys.  90, 015001 (2018).
\bibitem{kagome_arcs} Noam Morali, Rajib Batabyal, Pranab Kumar Nag, Enke Liu, Qiunan Xu, Yan Sun, Binghai Yan, Claudia Felser, Nurit Avraham, Haim Beidenkopf, 	Science Vol.365, 1286--1291 (2019). doi:10.1126/science.aav2334 
\bibitem{asymmr} S. Albarakati, C. Tan, Z. Chen, J. G. Partridge, G. Zheng, L. Farrar, E. L. H. Mayes, M. R. Field, C. Lee, Y. Wang, Y. Xiong, M. Tian, F. Xiang, A. R. Hamilton, O. A. Tretiakov, D. Culcer, Y. Zhao, and Y. Wang, Sci. Adv. 5, eaaw0409 (2019). https://doi.org/10.1126/sciadv.aaw0409

\bibitem{kagome} Enke Liu, Yan Sun, Nitesh Kumar, Lukas Muechler, Aili Sun, Lin Jiao, Shuo-Ying Yang, Defa Liu, Aiji Liang, Qiunan Xu, Johannes Kroder, Vicky S\"uss, Horst Borrmann, Chandra Shekhar, Zhaosheng Wang, Chuanying Xi, Wenhong Wang, Walter Schnelle, Steffen Wirth, Yulin Chen, Sebastian T. B. Goennenwein, and Claudia Felser, Nature Physics 14, 1125 (2018)
\bibitem{kagome1} Qi Wang, Yuanfeng Xu, Rui Lou, Zhonghao Liu, Man Li, Yaobo Huang, Dawei Shen, Hongming Weng, Shancai Wang and Hechang Lei, Nature Communications 9,  3681 (2018)


\bibitem{CoSnS_skyrmion} Akira Sugawara, Tetsuya Akashi, Mohamed A. Kassem, Yoshikazu Tabata, Takeshi Waki, and Hiroyuki Nakamura, Physical Review Materials 3, 104421 (2019).
\bibitem{CrGeTe} Myung-Geun Han, Joseph A. Garlow, Yu Liu, Huiqin Zhang, Jun Li, Donald DiMarzio, Mark W. Knight, Cedomir Petrovic, Deep Jariwala and Yimei Zhu, Nano Lett., 19, 11, 7859 (2019).
\bibitem{FGT_skyrmion} Bei Ding, Zefang Li, Guizhou Xu, Hang Li, Zhipeng Hou, Enke Liu, Xuekui Xi, Feng Xu, Yuan Yao, and Wenhong Wang, Nano Lett., 20, 868--873 (2020).




\bibitem{Orlova}A.A. Avakyants, N.N. Orlova, A.V. Timonina, N.N. Kolesnikov, E.V. Deviatov, Journal of Magnetism and Magnetic Materials, Vol. 573, p. 170668 (2023) https://doi.org/10.1016/j.jmmm.2023.170668

\bibitem{invhyst} M. Charilaou, C. Bordel, and F. Hellman, Appl. Phys. Lett. 104, 212405 (2014).
\bibitem{Exb} J. Nogu\'es, Ivan K. Schuller, Journal of Magnetism and Magnetic Materials, 192, 203 (1999).
\bibitem{Invh} M. J. O'Shea and A.-L. Al-Sharif, Journal of Applied Physics 75, 6673 (1994).
\bibitem{Eb_Invh_oxfilm} Mohammad Saghayezhian, Zhen Wang, Hangwen Guo, Rongying Jin, Yimei Zhu, Jiandi Zhang and E. W. Plummer Physical Review Research 1, 033160 (2019).
\bibitem{bow-tie} Felipe Tejo, Denilson Toneto, Sim\'on Oyarz\'un, Jos\'e Hermosilla, Caroline S. Danna, Juan L. Palma, Ricardo B. da Silva, Lucio S. Dorneles, and Juliano C. Denardin, ACS Appl. Mater. Interfaces, 12, 47, 53454 (2020).



\bibitem{MCEmater1} Altifani Rizky Hayyu, Stanislaw Baran and Andrzej Szytula, (2024) arXiv:2402.17912v1 
\bibitem{MCEmater2} Julia Lyubina, Journal of Physics D: Applied Physics  Vol. 50, 053002 (2017)
\bibitem{ref devices} V. Franco, J.S. Blazquez, J.J. Ipus, J.Y. Law, L.M. Moreno-Ramirez, A. Conde, Progress in Materials Science, Vol. 93, 112--232 (2018) 

\bibitem{multical1} Melvin M. Vopson, Yuri K. Fetisov, Ian Hepburn, Magnetochemistry 7, 12, 154 (2021) https://doi.org/10.3390/magnetochemistry7120154
\bibitem{multical2} Jia-Zheng Hao, Feng-Xia Hu , Zi-Bing Yu, Fei-Ran Shen, Hou-Bo Zhou, Yi-Hong Gao, Kai-Ming Qiao, Jia Li, Cheng Zhang, Wen-Hui Liang, Jing Wang, Jun He, Ji-Rong Sun and Bao-Gen Shen, Chin. Phys. B Vol. 29, No. 4  047504 (2020).
\bibitem{multical3} X. Wan, A. M. Turner, A. Vishwanath, S. Y. Savrasov,  Phys. Rev. B 83, 205101
(2011).
\bibitem{GMCE1} H. Chouaibi, S. Mansouri, S. Aitjmal, M. Balli, O. Chdil, M. Abbasi Eskandari, S. H. Bukhari, P. Fournier (2024), 	arXiv:2401.01431
\bibitem{GMCE2} Pedro Baptista de Castro, Kensei Terashima, Takafumi D. Yamamoto, Suguru Iwasaki, Ryo Matsumoto, Shintaro Adachi, Yoshito Saito, Hiroyuki Takeya and Yoshihiko Takano Science And Technology Of Advanced Materials, 21, NO. 1, 849 (2020)
 https://doi.org/10.1080/14686996.2020.1856629 

\bibitem{deltaS1/RMCE} Anzar Ali, Shama, Yogesh Singh, J. Appl. Phys. 126, 155107 (2019). https://doi.org/10.1063/1.5120005
\bibitem{deltaS2} Qi Shia, Xiao Zhanga, En Yanga, Jin Yana, Xiaoyun Yua, Chang Suna, Si Lib, Zhengwei Chen, Results in Physics, Vol. 11, pp. 1004--1007 (2018).  https://doi.org/10.1016/j.rinp.2018.10.027
\bibitem{deltaS3} Jiyu Hu, Xucai Kan, Zheng Chen, Ganhong Zheng, Yongqing Ma, Journal of the American Ceramic Society, 105, pp 4827--4839 (2022). DOI: 10.1111/jace.18465.


\bibitem{Tc1} W. Schnelle, A. Leithe-Jasper, H. Rosner, F. M. Schappacher, R. P\"ottgen, F. Pielnhofer, and R. Weihrich, Phys. Rev. B Vol. 88, 144404 (2013)
\bibitem{Tc2} Mohamed A. Kassem, Yoshikazu Tabata, Takeshi Waki and Hiroyuki Nakamura
Phys. Rev. B Vol. 96, 014429 (2017)
\bibitem{technique} V. Franco, Lake Shore Cryotronics, Inc. | t. 614.891.2243 | f. 614.818.1600| www.lakeshore.com


\bibitem{DeltaS} Tishin, A.M. and Spichkin, Y.I. (2003). The Magnetocaloric Effect and its Applications (1st ed.). CRC Press. https://doi.org/10.1201/9781420033373
\bibitem{DeltaTad} M. F\"olde\'aki, R. Chahine; T. K. Bose J. Appl. Phys. 77, 3528--3537 (1995), doi.org/10.1063/1.358648
\bibitem{Cp}	W. Schnelle, A. Leithe-Jasper, and H. Rosner, Physical Review B, Vol. 88, 144404 (2013)

\bibitem{CoSnS_Tdepend1} Z. Guguchia, J. A. T. Verezhak, D. J. Gawryluk, S. S. Tsirkin, J.-X. Yin, I. Belopolski, H. Zhou, G. Simutis, S.-S. Zhang, T. A. Cochran, G. Chang, E. Pomjakushina, L. Keller, Z. Skrzeczkowska, Q. Wang, H. C. Lei, R. Khasanov, A. Amato, S. Jia, T. Neupert, H. Luetkens and M. Z. Hasan, Nature Communications, 11, 559 (2020).
\bibitem{CoSnS_Tdepend2} H.C. Wu, P.J. Sun, D.J. Hsieh, H.J. Chen, D. Chandrasekhar Kakarla, L.Z. Deng, C.W. Chu, H.D. Yang, Materials Today Physics, 12, 100189 (2020).
\bibitem{CoSnS_spin glass} Ella Lachman, Ryan A. Murphy, Nikola Maksimovic, Robert Kealhofer, Shannon Haley, Ross D. McDonald, Jeffrey R. Long and James G. Analytis, Nature Communications, 11,  560 (2020).

\bibitem{CoSnS_Ws1} Noam Morali, Rajib Batabyal, Pranab Kumar Nag, Enke Liu, Qiunan Xu,Yan Sun, Binghai Yan, Claudia Felser, Nurit Avraham, Haim Beidenkopf, Science 365, 1286 (2019).
\bibitem{CoSnS_Ws2} Qi Wang, Yuanfeng Xu, Rui Lou, Zhonghao Liu, Man Li, Yaobo Huang, Dawei Shen, Hongming Weng, Shancai Wang and Hechang Lei, Nature Communications, 9, 3681 (2018).



\bibitem{Co/Pd} Robert Streubel, Luyang Han, Mi-Young Im, Florian Kronast, Ulrich K. R$\ddot{o}$ler, Florin Radu, Radu Abrudan, Gungun Lin, Oliver G. Schmidt, Peter Fischer and Denys Makarov, Scientific Reports, 5, 8787 (2015).
\bibitem{Pt/Co/Ta} Senfu Zhang, Junwei Zhang, Yan Wen, Eugene M. Chudnovsky and Xixiang Zhang, Communications Physics, 1, 36 (2018).
\bibitem{Pt/Co/Ta1} Senfu Zhang, Junwei Zhang, Qiang Zhang, Craig Barton, Volker Neu, Yuelei Zhao, Zhipeng Hou, Yan Wen, Chen Gong, Olga Kazakova, Wenhong Wang, Yong Peng, Dmitry A. Garanin, Eugene M. Chudnovsky and Xixiang Zhang, Applied Physics Letters 112, 132405 (2018).
\bibitem{Pt/Co/Ta2} You Ba, Shihao Zhuang, Yike Zhang, Yutong Wang, Yang Gao, Hengan Zhou, Mingfeng Chen, Weideng Sun, Quan Liu, Guozhi Chai, Jing Ma, Ying Zhang, Huanfang Tian, Haifeng Du, Wanjun Jiang, Cewen Nan, Jia-Mian Hu and Yonggang Zhao, Nature Communications, 12, 322 (2021); https://doi.org/10.1038/s41467-020-20528-y.
\bibitem{Ir/Fe/Co/Pt} Anjan Soumyanarayanan, M. Raju, A. L. Gonzalez Oyarce, Anthony K. C. Tan, Mi-Young Im, A. P. Petrovi\'c, Pin Ho, K. H. Khoo, M. Tran, C. K. Gan, F. Ernult and C. Panagopoulos, Nature Mater, 16, 898–904 (2017). https://doi.org/10.1038/nmat4934.



 




\end{thebibliography}
\end{document}